\documentclass[aps,prd,twocolumn,amssymb,amsmath,floatfix,nofootinbib,superscriptaddress]{revtex4-1}
\usepackage{graphicx}
\usepackage{dcolumn}
\usepackage{bm}
\usepackage[usenames, dvipsnames]{color}
\usepackage[normalem]{ulem}
\usepackage{sidecap}
\usepackage{subcaption}

\usepackage[colorlinks,bookmarks]{hyperref}
\definecolor{linkblue}{rgb}{0,0,0.8}
\definecolor{linkgreen}{rgb}{0,0.5,0}

\hypersetup{pdfpagemode=UseNone, pdfstartview=FitH, linkcolor=linkblue, %
            citecolor=linkgreen, urlcolor=linkblue}

\graphicspath{{Figures/}}

\newcommand{\VSI}{Van Swinderen Institute for Particle Physics and Gravity,\\ University of Groningen,
Nijenborgh 4, 9747 AG Groningen, The Netherlands}

\begin{document}

\title{Initial conditions of the universe: Decaying tensor modes}

\author{Darsh Kodwani}
\email{darsh.kodwani@physics.ox.ac.uk}
\affiliation{University of Oxford, Department of Physics, Denys Wilkinson Building, Keble Road, Oxford, OX1 3RH, UK.}

\author{P. Daniel Meerburg}
\email{pdm@ast.cam.uk}
\affiliation{\VSI}

\author{Ue-Li Pen}
\email{pen@cita.utoronto.ca}
\affiliation{Canadian Institute of Theoretical Astrophysics, 60 St George St, Toronto, ON M5S 3H8, Canada.}
\affiliation{Canadian Institute for Advanced Research, CIFAR program in Gravitation and Cosmology.}
\affiliation{Dunlap Institute for Astronomy \& Astrophysics, University of Toronto, AB 120-50 St. George Street, Toronto, ON M5S 3H4, Canada.}
\affiliation{Perimeter Institute of Theoretical Physics, 31 Caroline Street North, Waterloo, ON N2L 2Y5, Canada.}

\author{Xin Wang}
\email{wangxin35@mail.sysu.edu.cn}
\affiliation{School of Physics and Astronomy, Sun Yat-sen University, 2 Daxue Road, Zhuhai, China}
\affiliation{Canadian Institute of Theoretical Astrophysics, 60 St George St, Toronto, ON M5S 3H8, Canada.}

\begin{abstract}
Many models of the early universe predict that there should be primordial tensor perturbations. 
These leave an imprint into the temperature and polarisation anisotropies of the cosmic microwave background (CMB). 
The differential equation describing the primordial tensor perturbations is a second order differential equation and thus has two solutions.
Canonically, the decaying solution of this equation in radiation domination is dropped as it diverges at early times and on superhorizon scales while it is then suppressed at late times. 
Furthermore, if there is an inflationary phase prior to the radiation domination phase, the amplitude of the decaying mode will also be highly suppressed as it enters the radiation phase, thus its effect will be negligible. 
In this study we remain agnostic to the early universe models describing pre-radiation domination physics and allow this mode to be present and see what effect it has on the CMB anisotropies. 
We find that the decaying mode, if normalised at the same time on subhorizon scales as the growing mode leaves an imprint on the CMB anisotropies that is identical to the growing mode. 
This is a new conceptual understanding as it means the decaying mode cannot be much more constrained than the growing mode on sub-horizon scales.
Contrary to expectation, on large scales both modes are poorly constrained for a scale invariant spectrum, and the apparent divergence of the decaying mode does not lead to a divergent physical observable.
Quantitatively, the decaying mode can be more constrained both from temperature and polarisation anisotropies.
We use a model independent, non-parametric, approach to constrain both of these primordial tensor perturbations using the temperature and polarisation anisotropies.
We find that both modes are best constrained at the reionisation and recombination bumps and crucially, at the reionisation bump the decaying mode can be distinguished from the growing mode. 

\end{abstract}

\maketitle


\section{Introduction}

The CMB is the dominant observational probe when it comes to constraining models of the early universe.
The temperature and polarisation anisotropies in the CMB have been observed by several experiments over the last few decades \cite{0067-0049-208-2-20, Planck1,  refId0}. 
One of the main goals of future CMB experiments is to measure the polarisation anisotropies to greater precision, especially on large scales. 
In particular, the detection of B mode polarisation in the CMB is a primary target \cite{Core1, LiteBird} as they could be a signature of primordial tensor perturbations, i.e gravitational waves, which are predicted by a large number of inflationary theories (see \cite{Finelli:2016cyd, Baumann:2008aq} and references therein).
To calculate the effect of primordial tensor perturbations on the CMB we parametrise the primordial amplitude of the perturbations using a power law power spectrum and convolve that with the transfer functions for $B$-mode polarisation using a Boltzmann solver such as CAMB\footnote{https://camb.info} or CLASS\footnote{http://class-code.net}.

One implicit assumption in current cosmological analysis that search for primordial gravitational waves is that the primordial perturbations only have a single solution, the so-called \emph{growing} solution/mode. However, since the perturbations in the early universe are described by a second order differential equation, another solution exists known as the \emph{decaying} solution/mode. 
In a recent study \cite{Kodwani:2019ynt} the effect of this second, orthogonal, mode, was considered for scalar perturbations in radiation domination. By explicitly keeping the decaying scalar mode and describing the primordial power as a set of independent bins in $k$, the effects of the decaying scalar mode on the CMB anisotropies was studied.
By constraining the amplitude of the primordial power spectrum for a broad range of band powers it was found that the decaying mode is equally well constrained as the growing mode on subhorizon scales, whereas on superhorizon scales there is a divergence in the decaying mode anisotropy spectrum which means they are more constrained then growing modes by several orders of magnitude.

The aim of this paper is to extend this analysis by calculating the effect of the decaying \emph{tensor} mode in radiation domination on the CMB. To our knowledge, this has not been considered before, but given the broad theoretical interest in inflationary tensor modes and potentially far-reaching theoretical implications in case of a detection, it is timely to explore the effect such a mode could have on the CMB in case it was produced in the early Universe.  
Specifically, while inflation predicts negligible decaying modes, such modes can be generated in bouncing universe scenarios and a detection could open a new window onto the novel physics that describes the beginning of our universe.
As was described in \cite{Kodwani:2019ynt, Amendola:2004rt}, the decaying modes are not constant, but evolve with time outside the horizon. Therefore we must specify the time at which we start the evolution of these modes. In the case of scalar perturbations one has to be careful about which gauge we use to define the time at which we evolve the modes as they will have different dependence on time in different gauges (i.e Newtonian and Synchronous gauges). 
Tensor perturbations have a unique description in any gauge at linear order and therefore do not suffer from these ambiguities. 

Canonically the form of the primordial power spectrum (PPS), $P_T(k)$, of tensor perturbations is assumed to come from the growing mode only and can be parametrised as a power law with an amplitude, $A_T$, and spectral index $n_T$
\begin{equation}
	P_T(k) = A_T \left( \frac{k}{k_*} \right)^{n_T} \label{tensor_pps}
\end{equation}
Here $k_*$ is the pivot scale for tensor perturbations.
We relax both of these assumptions, allowing the PPS to have a decaying mode solution while its power is described by a non-parametric binned form of the power spectrum. Similar approaches have recently been used to analyse the growing mode PPS for tensor perturbations \cite{Hiramatsu:2018nfa, Farhang:2018wgm, Campeti:2019ylm}. 

The paper is structured as follows: In section \ref{decay_tensor} we briefly describe the theoretical framework of primordial tensor perturbations and the form of decaying initial conditions. Section \ref{anal} describes the formalism we use to constrain the decaying tensor initial conditions using a Fisher matrix analysis and presents the results. 
Finally, we summarise in section \ref{summary}.

\section{Decaying tensor modes}\label{decay_tensor}

\subsection{Primordial perturbations}

In this section we briefly review the formalism of computing tensor perturbations in the early universe and work out the form of the decaying modes (more detailed introductions can be found in \cite{Kodama:1985bj}, \cite{MUKHANOV1992203}).
The tensor perturbations are uniquely defined by perturbing the flat Minkowski metric 
\begin{equation}
	ds^2 = a^2(\tau) \left( -d \tau^2 + (\delta_{ij} + h_{ij}) dx^i dx^j \right), \label{TT}
\end{equation}
where $h_{00} = h_{0i} = 0, |h_{ij}|\ll 1$ and the perturbations are transverse and traceless $h_{ij,i} = h^i_i = 0$.
To linear order, the transverse traceless perturbations in  Eq.~\eqref{TT} are gauge invariant. 
The tensor perturbations have two polarisation states denoted by ($+, \times$). 
The equation of motion for the perturbations is given by solving the Einstein equation. 
It is easiest to solve it in Fourier space and therefore we decompose the tensor perturbations into plane waves of each polarisation mode
\begin{equation}
	h_{ij}(\tau, \textbf{x} ) = \sum_{\lambda} \int \frac{d^3k}{(2 \pi)^3} h_\lambda(\tau, \textbf{k}) e^{i\textbf{k} \cdot \textbf{x}} \epsilon^\lambda_{ij}.
\end{equation}
Here we have defined $\epsilon^\lambda_{ij}$ to be the polarisation tensor and $\lambda \in \{ +, \times \}$.
To linear order the Einstein equations for metric perturbations is given by (in units of $c = \hbar = M_{\rm pl} = 1$) a Klein-Gordon equation for a massless scalar field, for each polarisation mode, with a source term given by the anisotropic stress term
\begin{equation}
	\ddot{h}_{k,\lambda} + 2\frac{\dot{a}}{a} \dot{h}_{k.\lambda} + k^2 h_{k,\lambda} = 2 a^2 \Pi_{ij}, \label{EOM}
\end{equation}
where the dot denotes derivatives w.r.t conformal time. 
$\Pi_{ij}$ is the anisotropic stress of the fluid with stress energy tensor $T_{ij} = p g_{ij} + a^2 \Pi_{ij}$. 
The anisotropic stress is typically generated by neutrinos free-streaming in the early universe after they decouple at $z \sim 10^9$. 
It has been shown in Ref~\cite{Weinberg:2003ur} that the effect of this anisotropic stress is to damp the effects of primordial tensor perturbations in the $B$-mode power spectrum. 
As anisotropic stress is generated by causal mechanisms it will not have an effect on superhorizon scales. 
If we look at the solutions to Eq.~\eqref{EOM}, during radiation domination, and in the absence of anisotropic stress we find
\begin{equation}
	h^{\text{rad}}_k(x) = A^{\text{rad}}_k j_0(x) + B^{\text{rad}}_k y_0(x).  \label{h_modes}
\end{equation}
Here $x = k\tau$, where $j_0(x)$ and $y_0(x)$ represent the spherical Bessel functions of the first and second kind of zero order respectively
\begin{equation}
	j_0(x) = \frac{\sin{x}}{x}, \hspace{3mm} y_0(x) = -\frac{\cos{x}}{x}.
\end{equation}
The $k$ index represents the fact the amplitude can be different for different $k$ modes.
The mode proportional to $A^{\text{rad}}_k/B^{\text{rad}}_k$ is the \emph{growing/decaying} mode.
The initial conditions are usually set when $x\ll1$ (i.e early times superhorizon scales) and in this limit the behaviour of these modes is 
\begin{equation}
	h_k^{\text{rad}}(x\ll1) = A^{\text{rad}}_k + \frac{B^{\text{rad}}_k}{x}. \label{h_ini}
\end{equation}
Moreover, if we look at solutions to Eq.~\eqref{EOM} in a matter domination phase, which starts at $z \approx 3000$, then we find
\begin{eqnarray}
	& & h^{\text{mat}}_k(x) = 3\left[ A^{\text{mat}}_k \frac{j_1(x)}{x} + B^{\text{mat}}_k\frac{y_1(x)}{x} \right], \nonumber \\
	& & j_1(x) = \frac{\sin{x}}{x^2} - \frac{\cos{x}}{x}, \hspace{3mm} y_1(x) = - \frac{\cos{x}}{x^2} - \frac{\sin{x}}{x}.
\end{eqnarray}
At late times, $j_1(x\gg 1) \rightarrow - \frac{\cos{x}}{x}$ and $y_1(x \gg 1) \rightarrow - \frac{\sin{x}}{x}$.  
So even if the decaying mode solution is ignored during radiation domination, it can still source two possible modes during matter domination. 
The key difference between these modes is that their phases during each era will be opposite, i.e the modes are orthogonal to each other. 
We have shown this schematically in Figure \ref{cmbpol}.\footnote{
Further discussion on this can be found in \cite{Turner:1993vb, Wang:1995kb, Pritchard:2004qp}.}
Our non-parametric approach allows us to analyse the effect of $k$ modes with different amplitudes and phases precisely by isolating the effect they have on the CMB anisotropies.
The two point correlation function of these modes is canonically defined as 
\begin{eqnarray}
    \sum_\lambda \langle |h_{\lambda, \textbf{k}}(\tau)|^2 \rangle & \equiv &  \frac{2 \pi^2}{2k^3} P_T(k) |T(\tau, k)|^2. \label{tensor_evo}
\end{eqnarray}
Note that we have assumed the expectation value of the $h_{\lambda, \textbf{k}}$ does not have any directional dependence (this is because we have assumed spatial isotropy, i.e. $SO(3)$ symmetry).
The correlation function is then separated into a time dependent and independent term. 
The time dependent term is often called the transfer function $T(\tau, k)$ and only tracks the \emph{time evolution of a particular tensor mode}. This is not the same as the transfer function for the CMB anisotropies which track the impact the perturbations have on the CMB photons. The growing mode transfer function is only time dependent on sub-horizon scales, while for the decaying mode it is time dependent on all scales. The time independent part is given by the PPS.

\subsection{Review of CMB anisotropies}

The tensor perturbations will leave an imprint on temperature and polarisation anisotropies in the CMB. 
We can gain some insight into their structure by exploring the computation of the anisotropies analytically. 
The structure of the anisotropies due to primordial tensors has been studied before in \cite{Turner:1993vb, Wang:1995kb, Zaldarriaga:1996xe, Pritchard:2004qp}. 
Here we briefly review it in the presence of decaying modes. 
The Gaussian anisotropies of the CMB can be completely described by the angular correlation function, $C_\ell$, which can be written as
\begin{equation}
	C_\ell^{XY} = 4 \pi \int^\infty_0  \frac{dk}{k} P(k) |\Delta^X_\ell(k) \Delta^Y_\ell(k) |.
\end{equation}
$X,Y \in \{T, E, B\}$ are the observables (temperature and two polarisation modes) that are computed from the CMB photons. $\Delta^X_\ell(k)$ is the transfer function corresponding to the observable one is interested in. 
The temperature and polarisation transfer functions for tensor perturbations are given by \cite{Zaldarriaga:1996xe, Pritchard:2004qp}
\begin{eqnarray}
	& & \Delta^T_\ell(k) = \sqrt{\frac{(\ell + 2)!}{(\ell - 2)!}} \int^{\tau_0}_{0} d \tau S_T(k, \tau) \mathcal{P}^{T}_\ell(x), \nonumber \\ 
	& & \Delta^E_\ell(k) = \int^{\tau_0}_0 \ d\tau S_P(k, \tau) \mathcal{P}^{E}_\ell(x) \nonumber \\
	& & \Delta^B_\ell(k) = \int^{\tau_0}_0 \ d \tau \ S_P(k, \tau) \mathcal{P}^{B}_\ell(x). \label{transfers}
\end{eqnarray}
$\tau_0$ is the conformal time today. 
The leading order source functions and projection factors for temperature and polarisation are given by 
\begin{eqnarray}
	& & S_T(k, \tau) = - \dot{h}(k, \tau) e^{- \kappa} + g(\tau) \Psi(k, \tau), \nonumber \\
	& & S_P(k, \tau) = -g(\tau) \Psi(k, \tau), \label{sources} \nonumber \\
	& & \mathcal{P}^{T}_\ell(x) = \frac{j_\ell(x)}{x^2}, \nonumber \\
	& & \mathcal{P}^{E}_\ell(x) = - j_\ell(x) + j_\ell''(x) + 2\frac{j_\ell(x)}{x^2} + 4\frac{j'_\ell(x)}{x}, \nonumber \\
	& & \mathcal{P}^{B}_\ell(x) = 2 j'_\ell(x) + 4 \frac{j_\ell(x)}{x}.
\end{eqnarray}
$\Psi(k, \tau)$ is the Newtonian gravitational potential. 
$\kappa$ is the integrated Thomson cross section between $\tau$ and $\tau_0$
\begin{equation}
	\kappa = \int^{\tau_0}_\tau d \tau \ a n_e x_e \sigma_T,
\end{equation}
where we have defined $x_e$ as the ionisation fraction, $n_e$ as the electron number density and $\sigma_T$ is the Thomson cross section. 
We have also defined the visibility function $g(\tau) \equiv \dot{\kappa} e^{-\kappa}$. 

\begin{figure}[h]
  \centering
  \includegraphics[width=\linewidth]{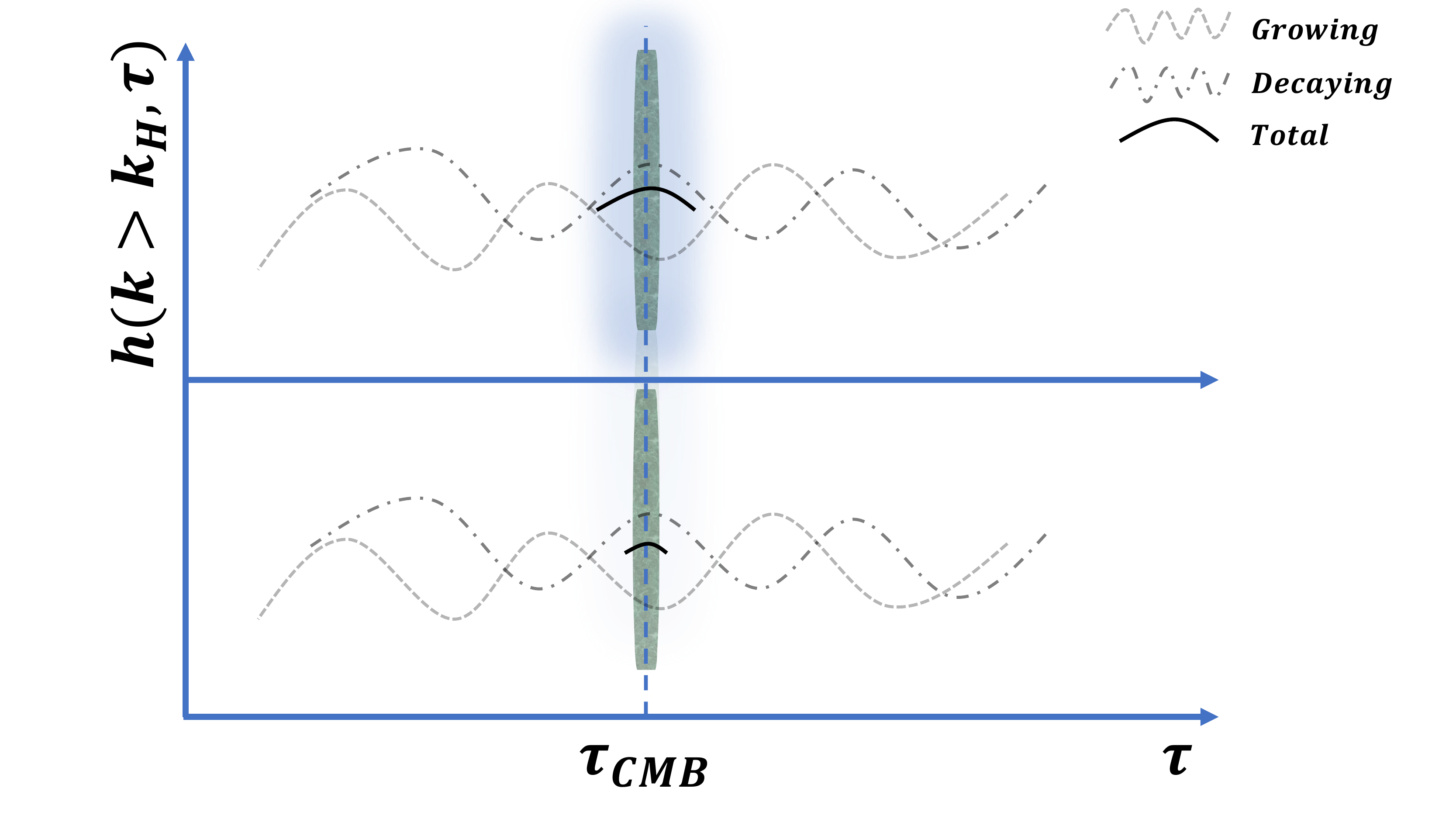}
  \caption{Schematic diagram showing the phase damping effect, as described in \cite{Pritchard:2004qp}, of primordial gravitational waves on the CMB photons. The top panel shows a situation where the last scattering screen is more diffuse then it is in the lower panel. A more diffuse screen will allow more of the gravitational wave amplitude to contribute to the production of polarisation anisotropy. }
  \label{cmbpol}
\end{figure}

From the expressions in Eq.~\eqref{transfers} and \eqref{sources} we can see that the temperature source function has two distinct features. 
The first term, proportional to $\dot{h}$, is a type of Integrated Sachs-Wolfe (ISW) effect. It describes the generation of anisotropies from the motion of photon geodesics in the presence of a time varying gravitational potential of the gravitational wave. 
The term proportional to the visibility function, $g$, will be localised to the screen of recombination, which is assumed to be almost instantaneous. 
This term is small and is almost always subdominant to the ISW term\footnote{See Figure 1 in \cite{Pritchard:2004qp} for instance.}. 
For the polarisation anisotropies, there is no ISW term. 
This is because the ISW effect changes the energy of the photon which is directly related to its temperature, not the polarisation.
The source of polarisation anisotropies will be strongly located at the surface of last scattering due to the scattering of free electrons from the local tensor quadrupole. 
Therefore, modes with $k \approx \frac{\ell}{\tau_0 - \tau_{\text{CMB}}}$ dominate the contribution to the polarisation $C_\ell$'s. 
Intuitively one would expect the anisotropies to be proportional to the width of the surface of last scattering as a larger width will lead to more polarisation being generated. This is because the finite width adds time for the generation of quadrupolar scattering, which is what fundamentally gives rise to the polarisation  anisotropy. 
In addition there will be more modes that can contribute to the anisotropy as the width increases. 
We show this schematically in the top panel of Figure \ref{cmbpol} which has a blurry CMB screen, whereas the bottom one has an almost instantaneous CMB.
Recombination is not the only screen present for the local tensor quadrupole to generate polarisation: 
Reionisation also provides another screen at which polarisation is generated and this happens at larger angular scales \cite{Zaldarriaga:1996xe}. 

If we assume the Bessel functions and their derivatives, which are usually referred to as projection factors, in Eqs.~\eqref{transfers} are approximately constant over the width of the CMB screen the $C_\ell$'s for polarisation can be calculated analytically by integrating over the source function and projection factors to get \cite{Pritchard:2004qp}
\begin{eqnarray}
	& & C^{EE/BB}_\ell \propto \int \frac{dk}{k} P_T(k) \mathcal{P}^{E/B}_\ell[k(\tau_0 - \tau_R)]^2 \dot{h}_k(\tau_{\text{CMB}})^2 \nonumber \\
	& & \times  \Delta \tau_{CMB}^2e^{- (\kappa \Delta \tau_{CMB})^2}.
\end{eqnarray}
Here we see that indeed the $C_\ell$'s are proportional to the finite time scale for recombination, $\Delta \tau_{CMB}$, and we also notice the $\dot{h}^2$ factor which will be sensitive to the initial conditions we choose. 
In particular, we see from Eq.~\eqref{h_ini} that the $C_\ell$'s will become large when $k\tau$ is small for the decaying modes (this is similar to what was seen in the case of decaying scalar perturbations \cite{Kodwani:2019ynt}) and the difference in phase of $h(\tau)$, or equivalently $\dot{h}(\tau)$, will only effect the amplitude of the $C_{\ell}$'s as the tensor perturbation is evaluated locally at $\tau_{CMB}$. 
This was described in \cite{Pritchard:2004qp} as phase damping because the overall effect of multiple phases is to damp the observed perturbations (also shown schematically in Figure \ref{cmbpol}). 

\section{Analysis}\label{anal}
\subsection{$C_\ell$ results}

\begin{figure*}
\centering
\begin{subfigure}{.5\textwidth}
  \centering
  \includegraphics[width=\linewidth]{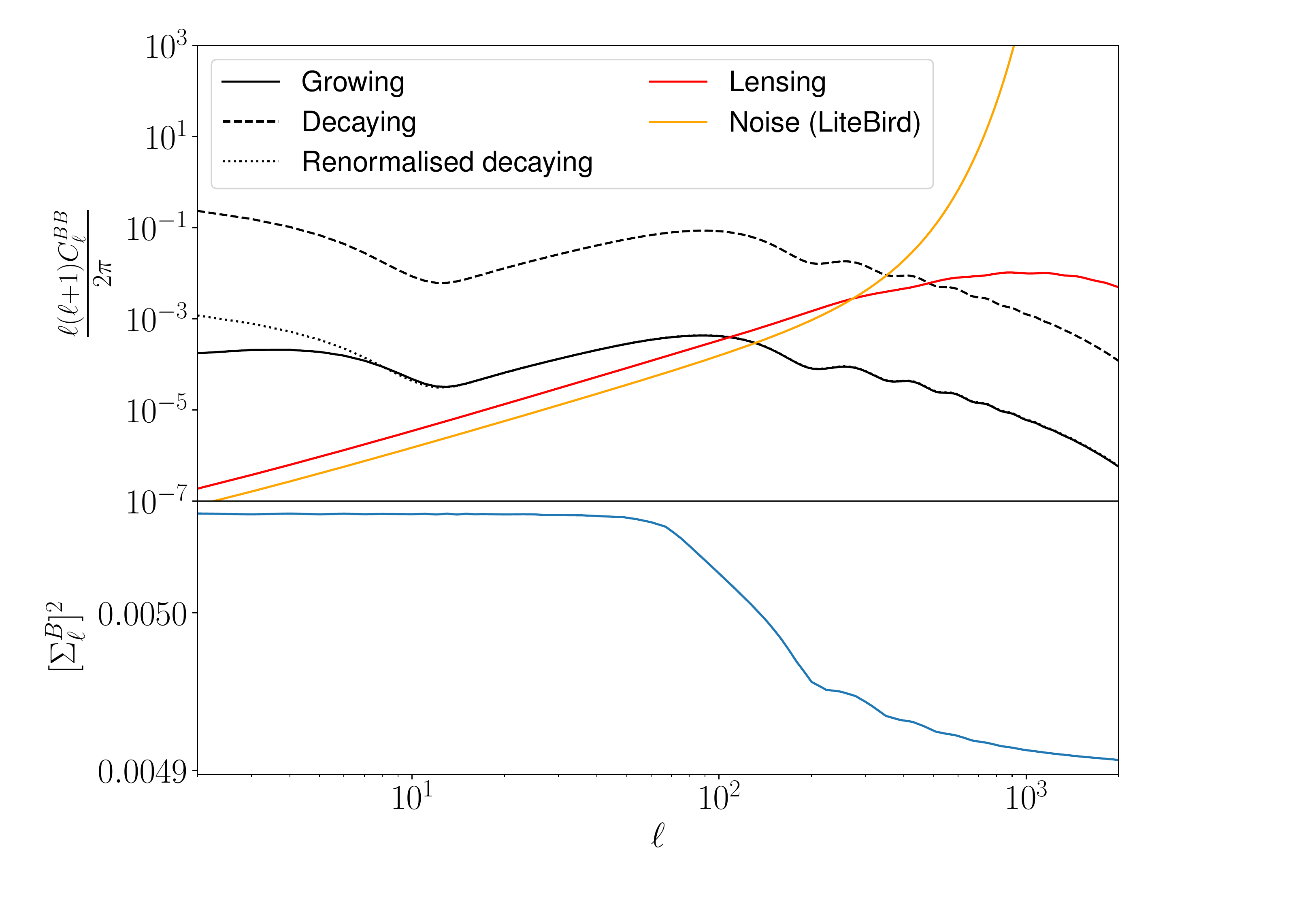}
  \caption{B mode}
  \label{clBB}
\end{subfigure}%
\begin{subfigure}{.5\textwidth}
  \centering
  \includegraphics[width=\linewidth]{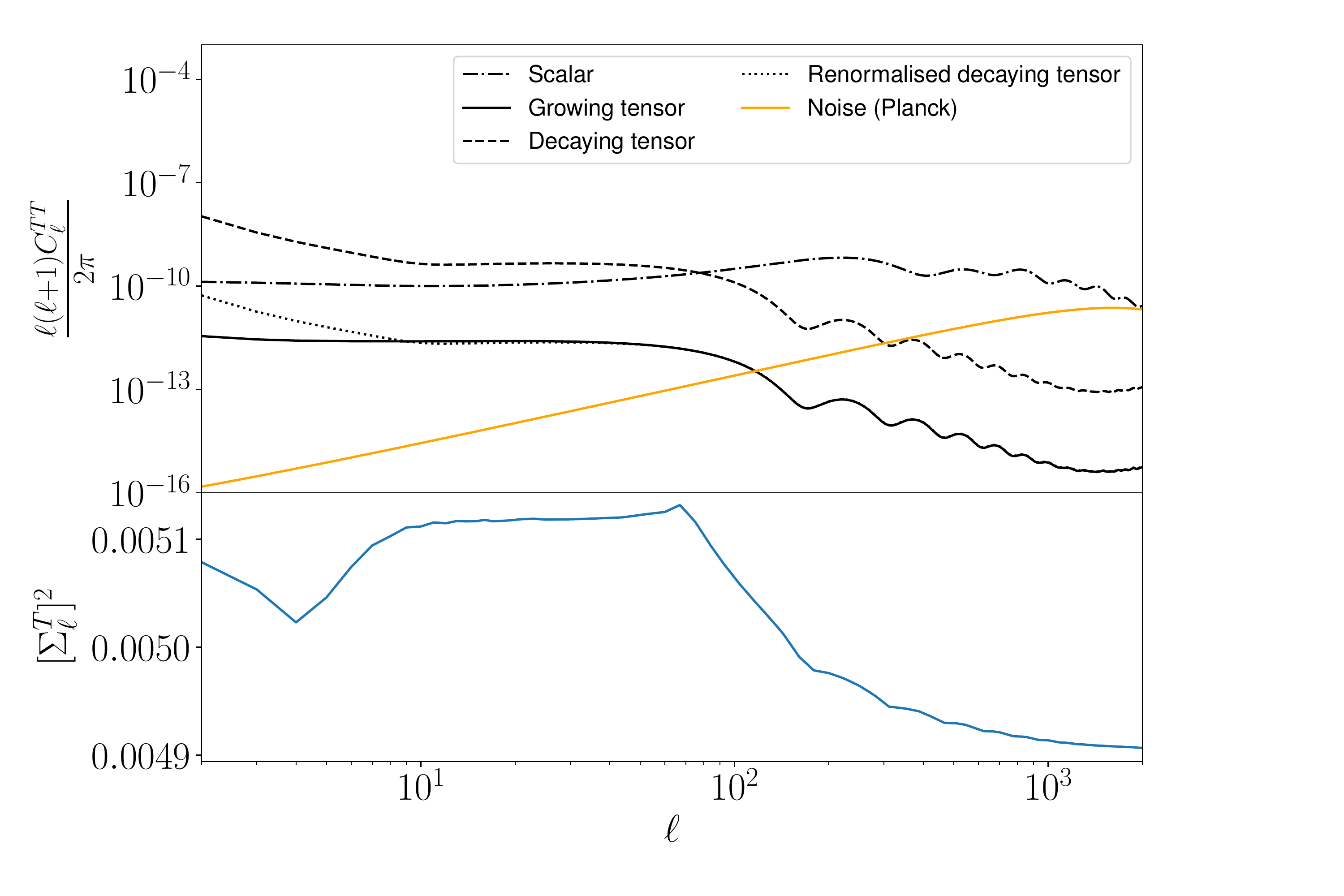}
  \caption{Temperature}
  \label{clTT}
\end{subfigure}

\begin{subfigure}{.5\textwidth}
  \centering
  \includegraphics[width=\linewidth]{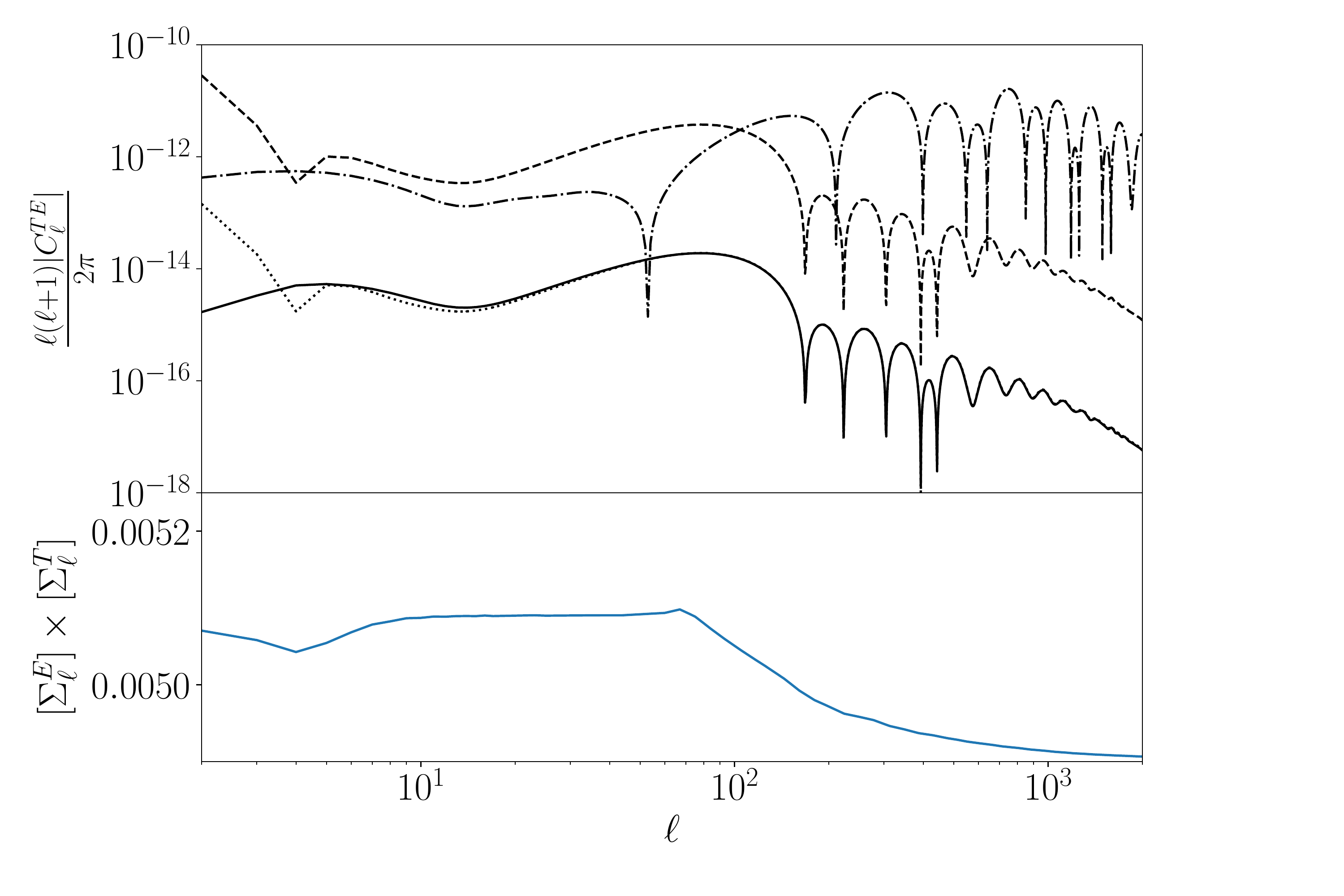}
  \caption{E mode + Temperature}
  \label{clTE}
\end{subfigure}%
\begin{subfigure}{.5\textwidth}
  \centering
  \includegraphics[width=\linewidth]{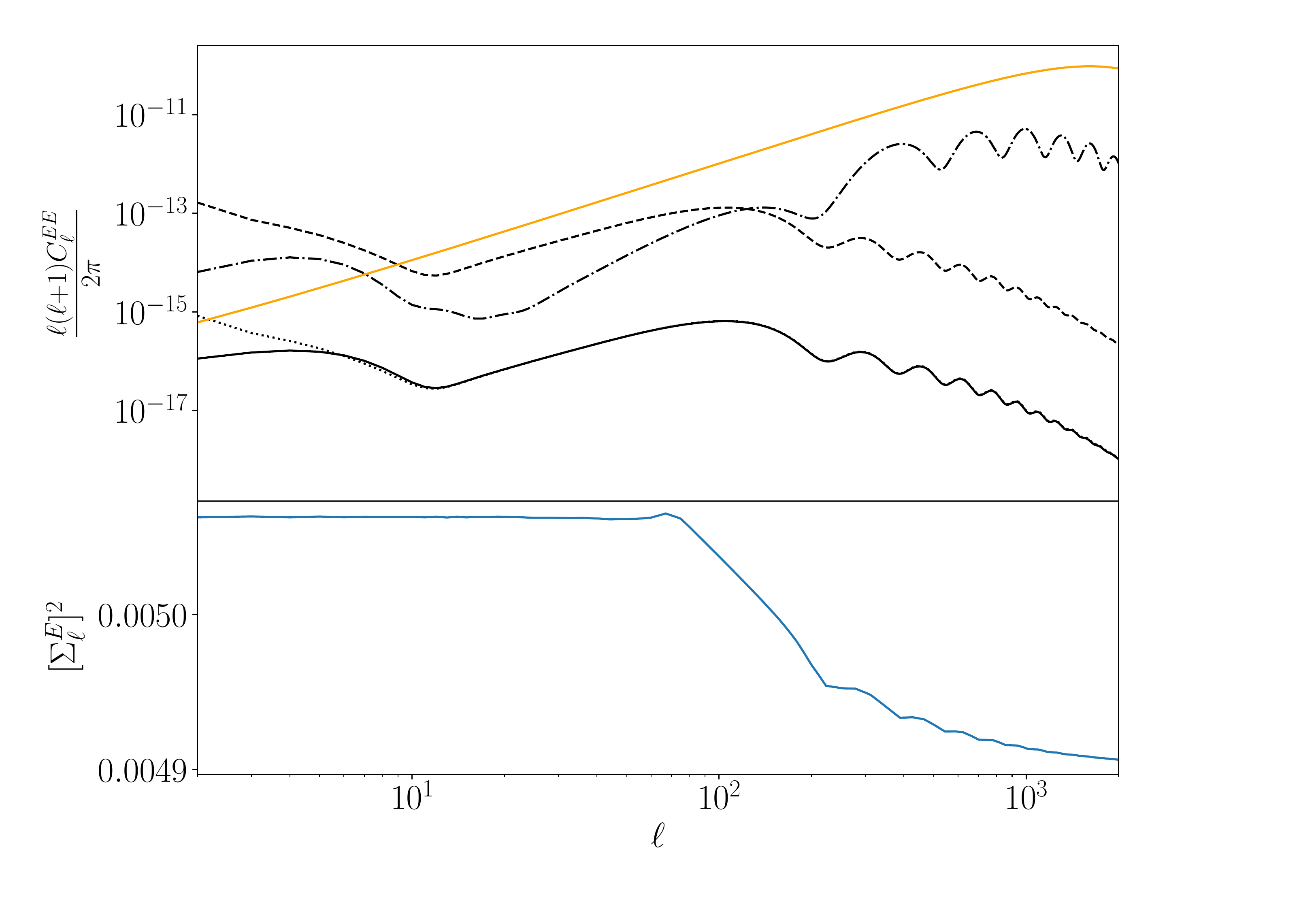}
  \caption{E mode}
  \label{clEE}
\end{subfigure}
\caption{Anisotropies for the growing and decaying tensor modes. 
The decaying mode is shown when it is normalised on superhorizon scales and when it is renormalised on subhorizon scales.
The renormalisation function for each observable is shown in the bottom panel of the plots.
The temperature and E mode polarisation spectra also show the contribution from the fiducial scalar perturbations in the standard $\Lambda$CDM cosmology with cosmological parameters given in Table \ref{cosmo_params}. Figures (\ref{clTE}, \ref{clEE}) have the same legend as Figure \ref{clTT}. The noise curves shown are for the LiteBird/Planck experiment for B/(T,E) modes which are defined in Eq.~\eqref{LiteBird_noise}/\eqref{planck_noise}.}
\label{decaying_tensor_cls}
\end{figure*}

To constrain tensor initial conditions we will follow the standard convention of using the scalar-to-tensor ratio $r = \frac{A_T}{A_S}$ as the variable to quantify the amplitude of the tensor perturbations. 
We implement the general initial condition in Eq \eqref{h_ini} in the CLASS Boltzmann code and plot the B mode power spectra in Figure \ref{clBB}. We show the growing mode with $r = 0.05$ and the growing mode spectral index defined by the single field slow roll inflation consistency relation $n_T^{(G)} = -\frac{1}{8} r^{(G)}$.
For the decaying mode we do not assume this relation (as we do not expect decaying modes from inflation to be detectable \cite{dePutter:2019xxv}) and set both the amplitude and index of the decaying mode independently.
In Figure \ref{clBB} we show the decaying mode B mode power spectra with the same amplitude as growing modes, $r^{(D)} = r^{(G)}$ and a scale invariant power spectrum, $n_T^{(D)} = 0$.
Furthermore, as the amplitude of the decaying mode is time dependent, one needs to define a normalisation time of when the $r^{(D)}$ is set.
In Figure \ref{decaying_tensor_cls} we plot the $C_\ell$'s for two different normalisations. 
First, when the amplitudes of the modes are set using the CLASS approximation schemes as described in \cite{CLASSII}. 
Under this scheme the amplitude of the modes on superhorizon scales is set at $\sim 0.01 \tau_{CMB}$.
The other normalisation procedure is the same as the one described in \cite{Kodwani:2019ynt}, where the tensor modes are renormalised such that the transfer function of both decaying and growing modes is the same on sub-horizon scales.
The renormalised transfer function for the decaying mode, $\tilde{\Delta}^{(D)}_\ell$, is defined as 
\begin{eqnarray}
    \tilde{\Delta}^{(D)}_\ell & = & \Delta^{(D)}_\ell \Sigma_\ell \nonumber \\
    \Sigma_\ell & \equiv & \frac{\int^{k_{\text{max}}}_{k_{\text{horizon}}} dk \ \Delta_\ell^{(G)}(k)} {\int^{k_{\text{max}}}_{k_{\text{horizon}}} dk \ \Delta^{(D)}_\ell(k)},
\end{eqnarray}
where $k_{\text{horizon}} = 3 \times 10^{-3} \ \text{Mpc}^{-1}$, $k_{\text{max}} = 2 \times 10^{-1} \ \text{Mpc}^{-1}$.
The renormalisation functions for each of the observables are shown in the bottom panels in Figure \ref{decaying_tensor_cls}.
The rest of the cosmological parameters are given in Table \ref{cosmo_params}.

\begin{center}
\begin{table}[h!]
\begin{tabular}{ p{1.5cm}  p{2cm}}
\hline
\hline
$A_s$ & 2.15 $\times 10^{-9}$  \\
$h$ & 0.67556  \\
$\Omega_bh^2$ & 0.022032  \\
$\Omega_{cdm}h^2$ & 0.12038 \\
$k_{*}$ & 0.002 Mpc$^{-1}$ \\
$n_s $ & 0.9619 \\
$N_{eff}$ & 3.046 \\
$r^{(G)}$ & 0.05 \\
$n_T^{(G)}$ & - $\frac{1}{8}r^{(G)}$ \\
\hline
$\ell_{max}$ & 2500 \\
$f_{sky}$ & 1 \\
\hline
\hline
\end{tabular}
\caption{Fiducial cosmological and systematic parameters}\label{cosmo_params}
\end{table}
\end{center}

In Figure \ref{decaying_tensor_cls} we see that when the decaying mode is normalised on superhorizon scales the anisotropies are larger than the growing mode ones. 
Furthermore, the decaying mode anisotropies for $TT$ and $EE$ can be even larger than the ones generated by \emph{scalar perturbations}\footnote{Here we are referring to the growing mode scalar perturbations, but the decaying scalar mode has a similar amplitude. See \cite{Kodwani:2019ynt} for further discussion on this.}. 
For temperature we see in Figure \ref{clTT} that the decaying mode anisotropy is greater than scalar perturbations for $ \ell \lesssim 90$. 
For E mode polarisation, seen in Figure \ref{clEE}, the decaying tensor mode contribution is always larger then the scalar contribution.
This means that if the modes are sourced at very early times on superhorizon scales they could already be constrained by the temperature and E mode polarisation anisotropies from Planck and WMAP.

Next, if we look at the anisotropies for the decaying mode when they are renormalised on subhorizon scales, we see that the shape of the $C_\ell$'s is the same, but the amplitude is smaller than the decaying mode sourced on superhorizon scales by a factor of $\sim 10^{4}$, which is as expected due to the superhorizon modes being sourced at $\sim 0.01 \ \tau_{CMB}$.
In this case the decaying modes are \emph{indistinguishable} from the growing tensor modes, except for a rise in anisotropy on very large scales. 
This rise is seen because we only renormalise based on the sub-horizon amplitude and the superhorizon amplitude will generally be larger on the large angular scales. 
The physical reason behind this is that the local quadrupole generated by the tensor modes is responsible for generating the polarisation in the CMB. 
As the decaying mode varies with time on superhorizon scales, the amplitude of the mode (and hence the polarisation it generates) depends on which time the mode is sourced.
The amplitude and the time the mode is sourced are degenerate parameters when it comes to the generation of the CMB polarisation. 
To break this degeneracy one needs to be able to measure the decaying mode at least twice and thus it will be important to measure the signal from reionisation and recombination. 
In particular, since we normalise the modes at recombination, the decaying and growing modes leave an identical signal in the B mode spectrum at the recombination bump, as can be seen from Figure \ref{clBB}.
The decaying mode can be distinguished from the growing mode only by the reionisation bump where we see an increase in power from the decaying mode.
Indeed we can get an idea as to how well the decaying mode can be measured from the B mode power spectrum by comparing the difference between the growing and decaying mode power spectra at the reionisation scale. By assuming a cosmic variance limited experiment, we know the variance in the $C_\ell$'s is given by
\begin{equation}
    \sigma (\ell)^2 = \frac{2}{2\ell+1} (C_\ell^{\text{G}})^2,
\end{equation} 
where $C_\ell^{\text{G}}$ is the power spectra for the fiducial growing modes. 
The difference between the decaying and growing mode power spectra is
\begin{equation}
    \Delta C_\ell^2 \equiv (C_\ell^{G} - C_\ell^{D})^2.
\end{equation}
The $C_\ell^{D}$ is the decaying mode power spectrum with parameters give in Table \ref{cosmo_params}.
At $\ell = 2$, where the signal is largest from the decaying mode we see that $\frac{\Delta C^2_{\ell = 2}}{\sigma(\ell=2)^2} \approx 80$.
This means the decaying mode can be measured at a statistically significant level in a cosmic variance limited experiment by measuring the B mode polarisation signal from reionisation at $\ell = 2$. 
To get a complete result accounting for the full covariance between the polarisation and temperature anisotropies as well as the total sum over all the modes we compute the Fisher information matrix of the amplitude of the modes in the next section.

Before we move on to computing the Fisher information it is worth pointing out that the increase in power on superhorizon scales comes from the fact that the decaying mode has a $1/k\tau$ behaviour, which leads a divergent amplitude in the power spectrum.
We also see that the divergence in the polarisation spectra at the reionisation scale is smaller than the divergence in the temperature spectrum.
This is because of the fundamental difference between how temperature and polarisation anisotropies are generated by tensor perturbations: the temperature anisotropies are sourced continuously by tensor perturbations whereas the polarisation anisotropies are sourced at fixed screens as described above, thus more modes leave an imprint in the temperature power spectrum (and therefore increase the amplitude more).

We know that in addition to primordial gravitational waves sourcing B modes, lensing of the CMB photons can also generate B mode polarisation, which we call $C_\ell^{BB,(L)}$. 
This is given by \cite{Lewis:2006fu, Hiramatsu:2018nfa}
\begin{equation}
	C_\ell^{BB, (L)} = \frac{1}{2\ell+1} \sum_{\ell' \ell''} \left(\mathcal{S}^{(-)}_{\ell \ell' \ell''} \right)^2 C^{EE}_{\ell'} C^{\phi \phi}_{\ell''},
\end{equation}
where 
\begin{eqnarray}
	\mathcal{S}^{(-)}_{\ell \ell' \ell''} & \equiv & \left[ \frac{ (2\ell+1)(2\ell'+1)(2\ell''+1)}{16 \pi} \right]^\frac{1}{2} \times \begin{pmatrix} \ell & \ell' & \ell'' \\ 2 & -2 & 0 \end{pmatrix}\nonumber \\
	& \times & \left[- \ell(\ell+1) + \ell'(\ell'+1) + \ell''(\ell''+1) \right] ,
\end{eqnarray}
with the term in the circular brackets being a Wigner 3j symbol.

In addition to these two physical effects generating a B mode, an experiment will also have a noise contribution for the B modes. 
We parametrise the effect of the noise by white noise with a smoothing beam assumed to be Gaussian \cite{Katayama_2011}\footnote{In principle there can also be a $\ell$ dependence in the noise but we do not address that in this study.}
\begin{equation}
	N_\ell^{BB} = \exp{\left( \frac{\ell^2 \sigma_b^2}{2} \right)} \left( \frac{ \pi}{10800} \frac{w_p^{-\frac{1}{2}}}{\mu \text{K arc min}} \right)^2 \mu \text{K}^2 \text{str}. \label{LiteBird_noise}
\end{equation} 
We assume a LiteBird\footnote{A satellite mission that will aim to measure the polarisation of the CMB \cite{LiteBird}.} like experiment with $\sigma_b = 3.7 \times 10^{-3}$ and $w_p = 1 \ \mu$K \cite{LiteBird}.
The various components of lensing and noise contributions, along with the B modes from primordial tensors are shown in Figure \ref{clBB}.
In the next section we investigate this further in a model independent, non-parametric way, by computing the Fisher information.

\subsection{Fisher results}

\begin{figure}
  \centering
  \includegraphics[width=\linewidth]{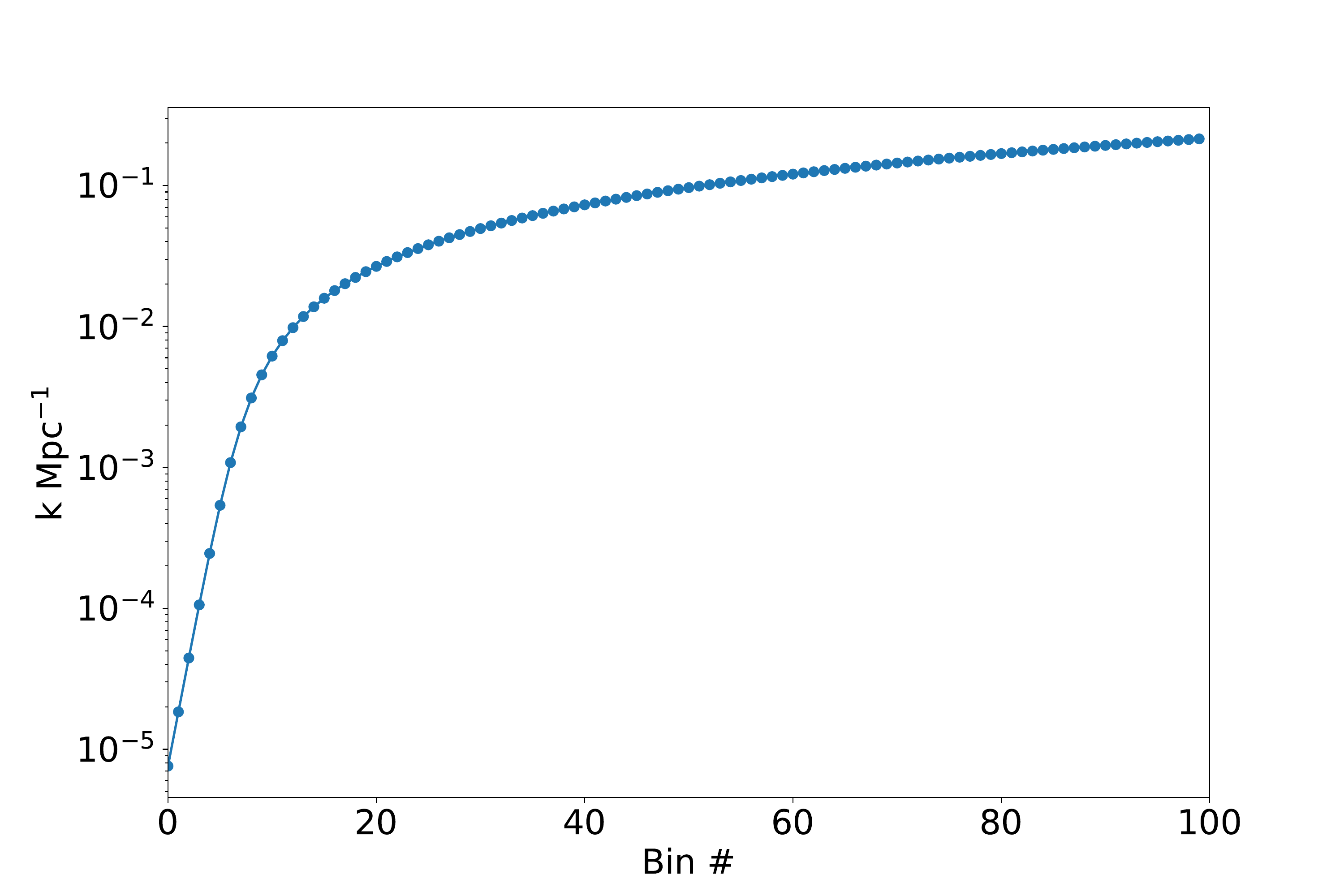}
  \caption{100 $k$ values use to compute the Fisher information.}
  \label{kbins}
\end{figure}

To obtain a model independent parameterisation of the PPS we model it as a set of 100 bins in $k$ around a fiducial PPS for the standard growing mode
\begin{equation}
	P_T(k,k_0,\epsilon) = \begin{cases} P_T(k)^{(G)} + \epsilon^{(G) \text{ or } (D)}_{k_0} & \mbox{if $k_0 = k$} \\ P_T(k)^{(G)} & \text{otherwise} \end{cases}. \label{pps}
\end{equation}
where $P_T(k)^{(G)}$ takes the form in Eq \eqref{tensor_pps}. $\epsilon_{k_0}^{(D) \text{ or } (G)}$ is the amplitude of additional power coming from the decaying or growing mode at the  scale $k_0$.
We treat the $\epsilon$'s in each $k$ bin as free parameters and constrain them using the Fisher information matrix, $F_{\alpha \beta}$. 
The $k$ bins we use are shown in Figure \ref{kbins}. 

For our Fisher analysis we focus solely on the decaying modes that are normalised on subhorizon scales as those are the physical modes we can observe at the time of decoupling. 
For modes that are sourced at early time and on superhorizon scales, the Fisher constraints can be scaled accordingly depending on what time the mode is sourced. 
For instance, if the mode is sourced at $0.01\tau_{CMB}$, the constraint will increase by $\sim 10^4$ for those scales due to the $1/k\tau$ behaviour of the decaying mode on superhorizon scales during radiation domination.
We assume a Gaussian likelihood with a parameter independent covariance matrix for the $C_\ell$'s and the corresponding Fisher matrix is 
\begin{equation}
	F_{\alpha \beta} = \frac{f_{sky}}{2} \sum_{\ell = 2}^{\ell_{max}} (2\ell + 1) \text{Tr} \left( \mathbb{C}^{-1}_\ell  \partial_\alpha \mathbb{C}_\ell \mathbb{C}^{-1}_{\ell} \partial_\beta \mathbb{C}_\ell \right)
\end{equation}
where
\begin{equation}
	\mathbb{C}_\ell \equiv \begin{pmatrix} \hat{C}_\ell^{TT} & C_\ell^{TE} & 0  \\ C^{ET}_\ell & \hat{C}^{EE}_\ell & 0 \\ 0 & 0 & \hat{C}^{BB}_\ell  \end{pmatrix}. \label{covmat}
\end{equation}
There are no correlations between E, T and B modes as long both polarisations of the tensor mode are equally generated (i.e there is no breaking of parity).
The $\hat{C}_\ell$ represents the theoretical $C_\ell$ (computed from a modified version of the CLASS Boltzmann code) plus noise contributions.
For each of these modes, these are defined by 
\begin{eqnarray}
	& & \hat{C}_\ell^{TT(EE)} \equiv C_\ell^{TT(EE)} + N_{\ell}^{TT(EE)} \nonumber \\
	& & \hat{C}_\ell^{BB} \equiv C_\ell^{BB} + N_\ell^{BB} + \lambda_{(L)}C_\ell^{BB, (L)}
\end{eqnarray}
where the B mode noise is defined in Eq~\eqref{LiteBird_noise}. 
We have introduced a lensing parameter $\lambda_{(L)}$ which denotes how much the lensing B modes contribute to the signal. $\lambda_{(L)} = 0$ corresponds to a situation where the lensing signal has been completely accounted for and removed from the signal.
The T and E mode noise is modelled by Gaussian random noise in 4 frequency channels given in the Planck blue book \cite{Planck:2006aa}
\begin{eqnarray}	
	N_\ell^{TT(EE)} & = & \left( (\sigma^2_{T(E)} B_\ell^2)_{100} + ( \sigma^2_{T(E)} B_\ell^2)_{143} \right. \nonumber \\
	& + & \left. (\sigma^2_{T(E)} B_\ell^2 )_{217} + (\sigma^2_{T(E)} B^2_\ell)_{353} \right)^{-1}. \label{planck_noise}
\end{eqnarray}
The window function for the beam is defined by $B_\ell^2 \equiv \exp{ \left( - \frac{\ell(\ell+1) \theta^2_{beam}}{8 \ln 2} \right) }$ and the variance for each frequency channel is $\sigma_{T(E)}$ for temperature/polarisation. 
The numerical values are given in Table \ref{Planck_noise_table} and the plot of the noise curves is shown in Figure \ref{clTT} and \ref{clEE}.
Once the Fisher information matrix is computed, the errors on the parameters is simply given by $(F^{-1}_{\alpha \alpha})^{\frac{1}{2}}$.

\begin{center}
\begin{table}[h!]
\begin{tabular}{ |p{1.5cm}p{2cm}p{2 cm}p{2 cm}|}
\hline
Frequency ($GHz$) & $\theta_{beam} (rad)$ &  $\sigma_T$($\mu K$ - rad) & $\sigma_E$($\mu K$ - rad)   \\
\hline
100 & 0.002763 & 0.001984 & 0.003174 \\
143 & 0.002065 & 0.001746 & 0.003333 \\
217 & 0.001454 & 0.003809 & 0.007785 \\
353 & 0.001454 & 0.011665 & 0.023647 \\
\hline
\end{tabular}
\caption{Planck noise parameters}\label{Planck_noise_table}
\end{table}
\end{center}

\begin{figure*}
\centering
\begin{subfigure}{.5\textwidth}
  \centering
  \includegraphics[width=\linewidth]{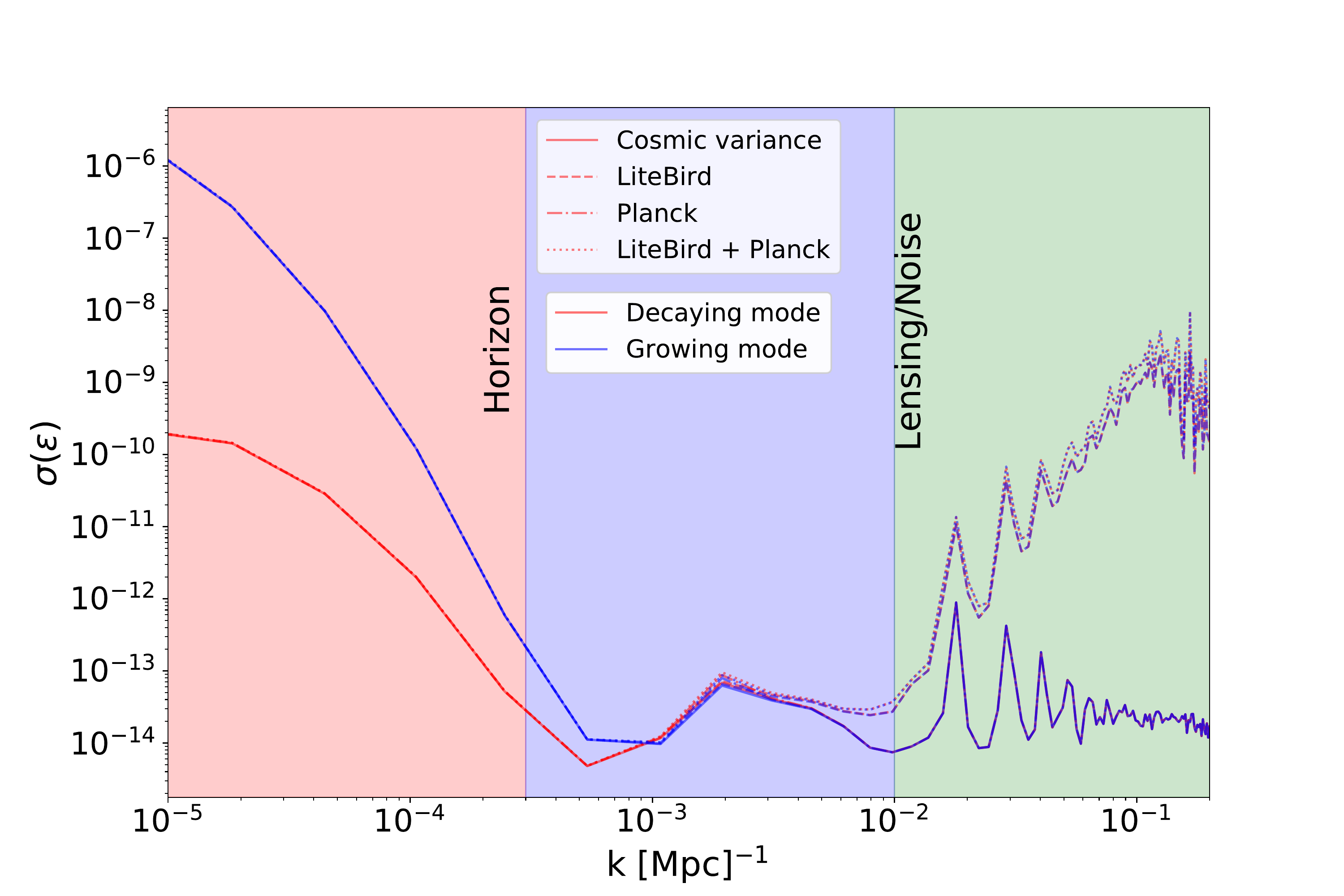}
  \caption{De-lensed B+E+T}
  \label{D_BTE}
\end{subfigure}%
\begin{subfigure}{.5\textwidth}
  \centering
  \includegraphics[width=\linewidth]{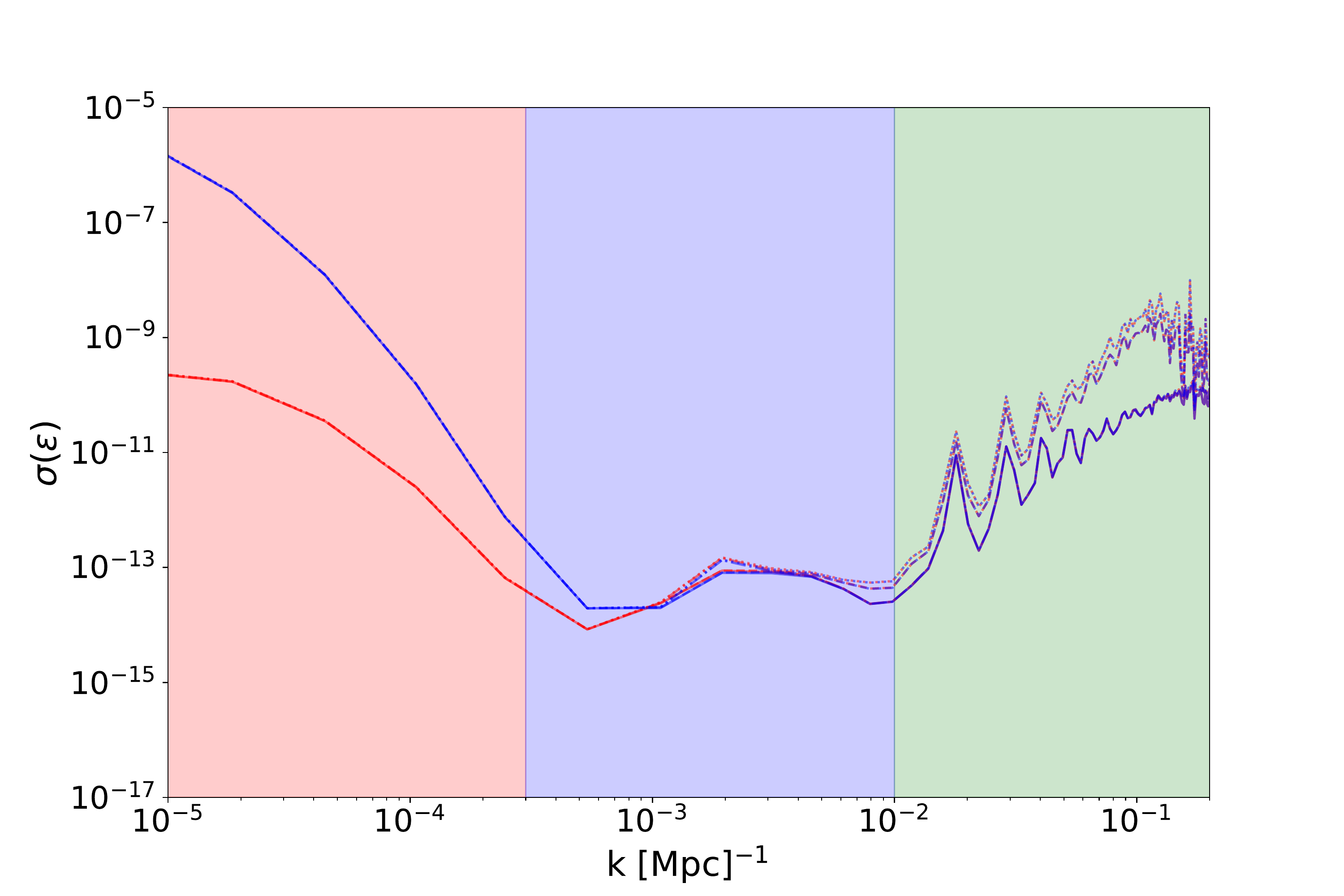}
  \caption{Lensed B+E+T}
  \label{L_BTE}
\end{subfigure}

\begin{subfigure}{.5\textwidth}
  \centering
  \includegraphics[width=\linewidth]{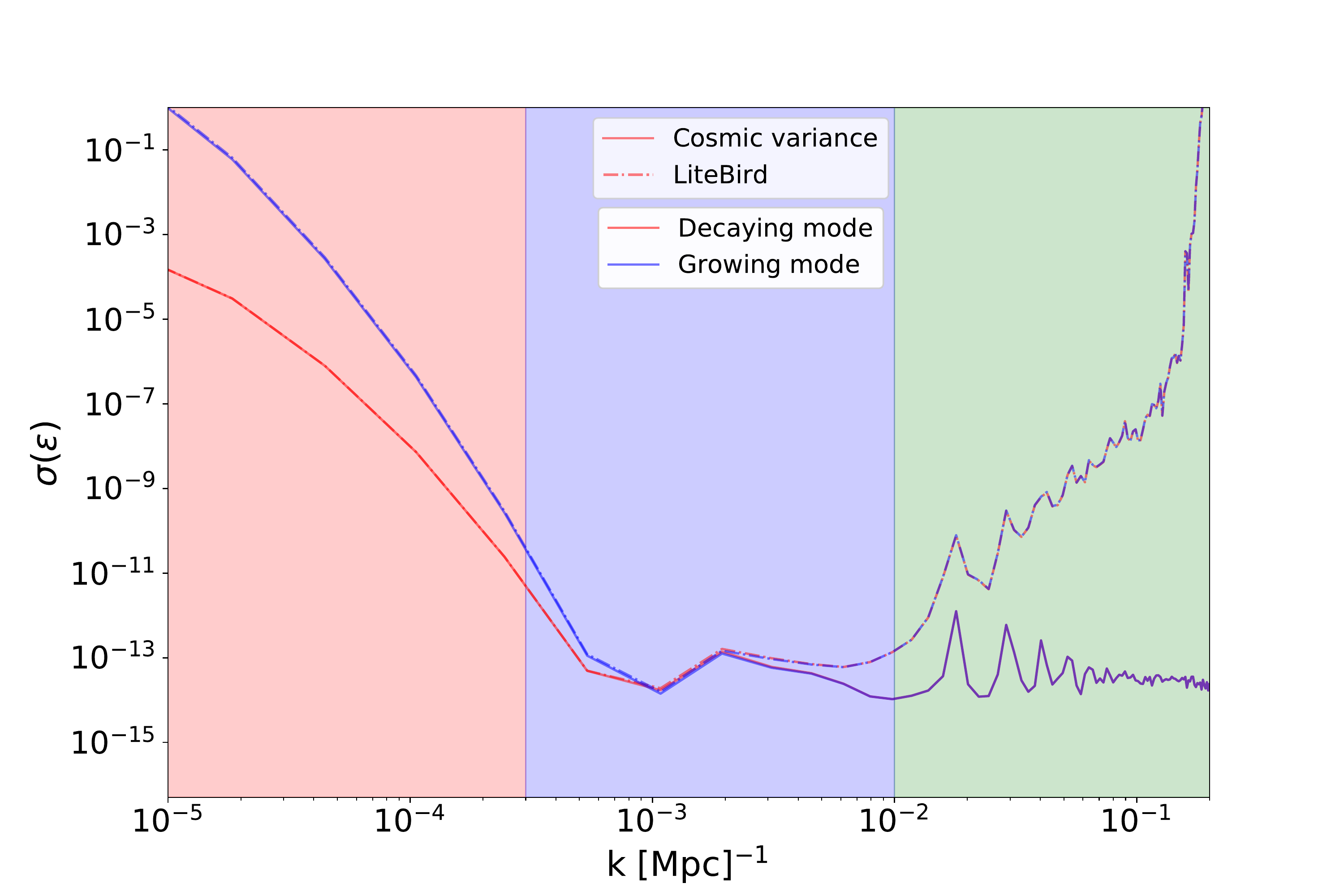}
  \caption{De-lensed B mode only}
  \label{D_B}
\end{subfigure}%
\begin{subfigure}{.5\textwidth}
  \centering
  \includegraphics[width=\linewidth]{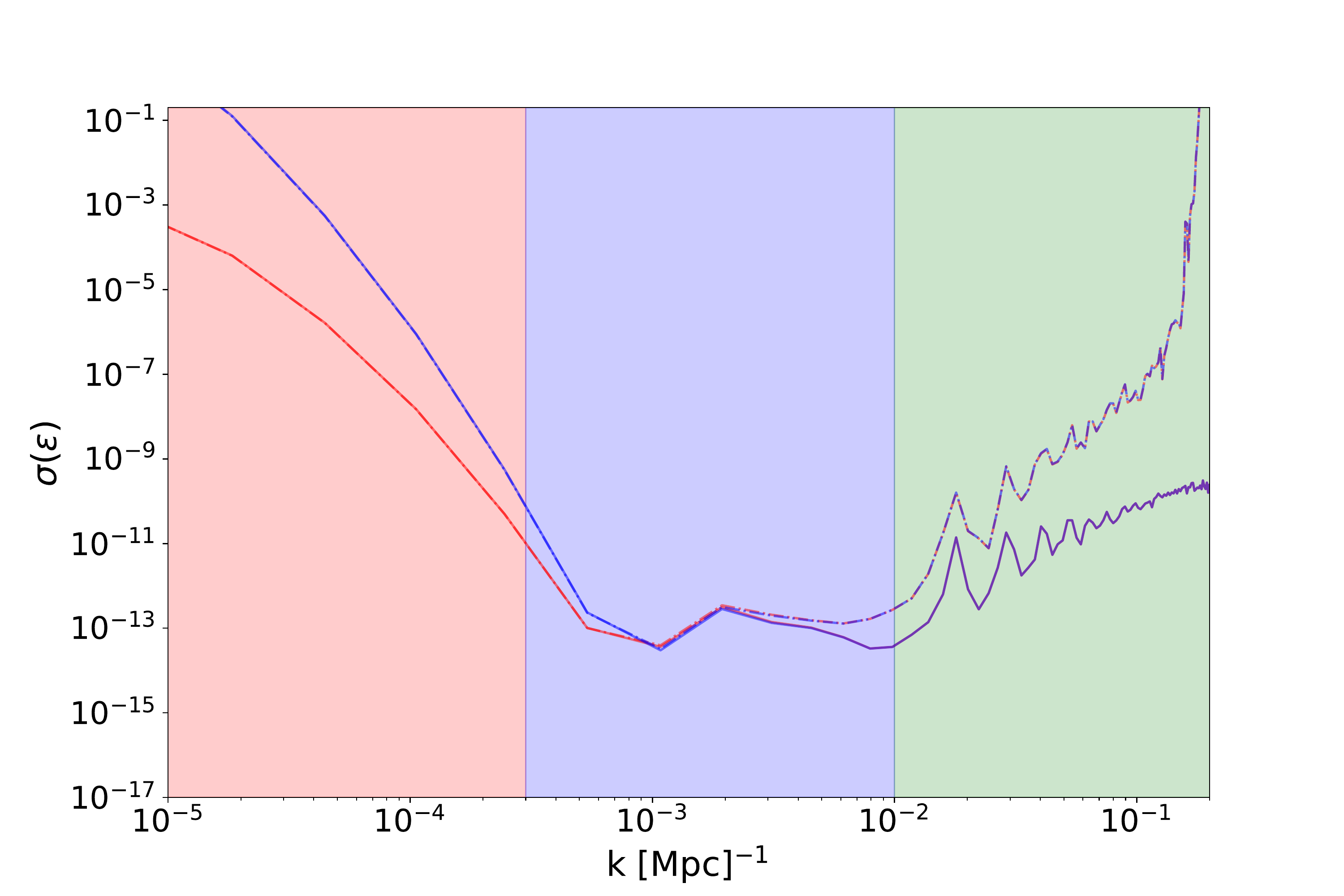}
  \caption{Lensed B mode only}
  \label{L_B}
\end{subfigure}
\caption{Errors for decaying and growing tensor modes. We have shown the errors for the four cases described in Table \ref{Resultsummary}. We have separated the noise contributions into the cosmic variance limited experiments, Planck noise for temperature and E mode polarisation and LiteBird for B mode polarisation.}
\label{fish_fig}
\end{figure*}

We show the errors on the PPS parameters $\epsilon_{k_0}$ in Eq.~\eqref{pps} for the growing and decaying tensor modes in Figure \ref{fish_fig}.
We focus on four cases which are summarised in Table \ref{Resultsummary}. 
The tracers used in the computation of the Fisher matrix are either B mode polarisation only, in which case $\mathbb{C}_\ell$  in Eq \eqref{covmat} is simply given by $\hat{C}_\ell^{BB}$, or B mode + E mode polarisation with temperature anisotropies as well. In this case we use the full $\mathbb{C}_\ell$ given in Eq \eqref{covmat}. This is denoted by T+E+B in Table \ref{Resultsummary}. For each of these cases, we consider the case when the modes are lensed/delensed with $(\lambda_{(L)} = 1) /(\lambda_{(L)} = 0) $. 

It is easiest to interpret the results in Figure \ref{fish_fig} by focusing on three different scales. 
First is the region shaded red which represents modes that are outside the horizon at the time the CMB is emitted (in fact there are scales that are larger than the observable size of the universe, thus one must be careful in how to interpret those constraints as we discuss in section \ref{summary}). 
This is also the region where cosmic variance dominates and thus the error bars increase substantially.
The region shaded in blue corresponds to scales which are subhorizon but on which the effect of lensing and noise (LiteBird experiment) for B modes is subdominant. 
Therefore the blue region is where most of the constraining power is. 
Finally, the green region is where the noise from LiteBird becomes very large and also the lensing contribution to B modes dominates over the primordial B mode signal. 

We see that the decaying mode is equally well constrained as the growing modes for all four cases we consider, except on scales below the recombination scale, $ k \lesssim 3 \times 10^{-4}$, where the decaying and growing mode amplitudes become distinguishable.
On superhorizon scales the constraint is $\sim 10^4$ larger for the decaying mode amplitude, as expected by the normalisation on subhorizon scales and the divergence of the decaying mode on superhorizon scales.
In the green region we see that the LiteBird noise dominates any signal and therefore the constraining power is reduced by 4-5 orders of magnitude. 
In the case of a cosmic variance limited experiment there is still the same amount of information in the green region as there is in the blue region when the CMB is delensed. If there is a lensing signal as well, the constraining power deteriorates by roughly 1-2 orders of magnitude.
When the temperature and E mode information is added we see that the errors on superhorizon scales, in the red region, are smaller by roughly 5 orders of magnitude. 
There are two reasons for this increase in constraining power. First, there is an increase in the $TT$ and $EE$ power spectra on superhorizon scales for the decaying mode. Second, the $TT$ and $EE$ $C_\ell$'s have different transfer functions to the $BB$, however the PPS for the decaying mode is the same for all of the observables. Therefore, the freedom in PPS is not able to compensate for the different transfer functions to the same extent when there are three observables.
 
The best constrained modes in all cases are at $k \approx 5 \times 10^{-4}$ Mpc$^{-1}$ and $\approx 7 \times 10^{-3 }$ Mpc$^{-1}$.
The physical reason behind this is that the polarisation is generated, and hence best constrained, when there is a anisotropic scattering of photons which happens at recombination and reionisation\footnote{This was also pointed out in this recent study \cite{Hiramatsu:2018nfa}.}. 
The recombination scale corresponds to a scale of $\ell \sim 80 $, which, in $k$ space corresponds to $k_{\text{recom}} \approx 6 \times 10^{-3}$ Mpc$^{-1}$.
Similarly the reionisation scale is given by $k_{\text{reion}} \approx 6 \times 10^{-4}$ Mpc$^{-1}$.

\begin{center}
\begin{table}[h!]
\begin{tabular}{ |p{1.5cm}p{2cm}p{2 cm} p{2 cm} |}
\hline
 & Tracer used & De-lensed & Result \\
\hline
\hline
Case 1 & B+E+T &  yes & figure \ref{D_BTE} \\
Case 2 & B+E+T &  no & figure \ref{L_BTE} \\
Case 3 & B &  yes & figure \ref{D_B} \\
Case 4 & B &  no & figure \ref{L_B} \\
\hline
\end{tabular}
\caption{Summary of different cases used to compute the errors on the PPS.}
\label{Resultsummary}
\end{table}
\end{center}

\section{Discussion \& future outlook}\label{summary}

In this paper we have analysed the effect a decaying tensor mode has on the CMB temperature and polarisation anisotropies.
The decaying modes evolve on superhorizon scales and thus the amplitude of these modes is degenerate with the time at which they are sourced.
We used a Fisher matrix formalism with a non-parametric binned PPS to understand the constraints on these modes.
If the decaying modes are sourced at very early times before decoupling, then they are highly constrained. 
If they are sourced on sub-horizon scales with same power as the growing mode, then there could be an ambiguity as to which mode generates the observed $B$-mode polarisation pattern. 
The amplitudes of both modes start to become distinguishable around the reionisation bump, which suggests it could be important to measure the $B$ modes on large scales $\ell \sim 5$.
If we only look on observable scales, i.e modes that are sub-horizon at the time of decoupling, the decaying and growing modes are constrained equally well. On super-horizon scales where the decaying mode becomes distinguishable from the growing mode it is more constrained. This is because it generates more power in the anisotropies due to its $1/k\tau$ scaling.
Decaying modes generated during inflation would be highly suppressed in radiation domination. Thus, if such modes are observed, it will be a unique signature of new physics on very high energies in the early universe. In particular, bouncing models could be a source of decaying modes \cite{Kodwani:2019ynt, Gielen:2016fdb}. 
There is a fundamental question that needs to be answered, however, in order to understand these modes. 
As the effect of the decaying mode is most apparent on super-horizon scales, it is worth asking how super-horizon tensor modes, specifically modes that are much larger than our current horizon, can or will effect our observable universe.
In the case of scalar perturbations it is possible the effect of these super-horizon modes will come from either a modification to overall background density, as is modelled in separate universe approached to cosmological perturbations \cite{Rigopoulos:2003ak}, or through the effects of spatial gradients \cite{Tanaka:2006zp}. 
For tensor modes, however, it is not clear what the dominant effect would be. For instance, it is possible that a large scale tensor mode modifies our patch of the universe to have an anisotropic metric, which for instance has been considered in the context of lensing in \cite{Adamek:2015mna}.
In this case the observable effect of the decaying tensor mode would actually be the presence of shear modes in the universe. More formal calculations of the shear modes can be found in \cite{Pontzen:2010eg, Ellis:1968vb, Pontzen:2009rx, 1982PhRvD..26.2951M}. 
Recent searches for shear modes in a general class of Bianchi models can be found in \cite{Saadeh:2016sak}. 
While shear modes are highly constrained, relating the decaying modes to the constraints on shear modes will require a gauge invariant description of matching super-horizon decaying tensor modes to the shear modes.
This would be an interesting endeavour and we leave that for future works.

\section*{Acknowledgments}
\noindent

This project has received funding from the European Research Council (ERC) under the European Union’s Horizon 2020 research and innovation programme (grant agreement No 693024).
P.~D.~M.\ acknowledges support from  the Netherlands organization for scientific
research (NWO) VIDI grant (dossier 639.042.730). We acknowledge helpful discussions with Latham Boyle, Pedro Ferreira and Neil Turok.

\bibliography{all_active}

\end{document}